\documentclass{article}

\usepackage{arxiv}

\usepackage[utf8]{inputenc} 
\usepackage{hyperref}       
\usepackage{url}            
\usepackage{booktabs}       
\usepackage{amsfonts}       
\usepackage{nicefrac}       
\usepackage{microtype}      
\usepackage{lipsum}
\usepackage{graphicx}
\usepackage{multirow}
\usepackage{subfigure}
\usepackage{natbib}

\title{Practice meets Principle: Tracking Software and Data Citations to Zenodo DOIs}

\author{
    Stephanie van de Sandt\\
    CERN\\
    Espl. des Particules 1\\
    1211 Meyrin, Switzerland\\
    \texttt{stephanie.van.de.sandt@cern.ch}
   \And
    Lars Holm Nielsen\\
    CERN\\
    Espl. des Particules 1\\
    1211 Meyrin, Switzerland\\
    \texttt{lars.holm.nielsen@cern.ch}
    \And
    Alexandros Ioannidis\\
    CERN\\
    Espl. des Particules 1\\
    1211 Meyrin, Switzerland
    \And
    Dr. August Muench\\
    American Astronomical Society\\
    1667 K Street NW\\
    Suite 800 Washington, DC 20006, USA
    \And
    Edwin Henneken\\ 
    Center for Astrophysics\\
    Harvard \& Smithsonian\\
    60 Garden Street\\
    Cambridge, MA 02138, USA
    \And
    Dr. Alberto Accomazzi\\
    Center for Astrophysics\\
    Harvard \& Smithsonian\\
    60 Garden Street\\
    Cambridge, MA 02138, USA
    \And
    Chiara Bigarella\\
    CERN\\
    Espl. des Particules 1\\
    1211 Meyrin, Switzerland
    \And
    Jose Benito Gonzalez Lopez\\
    CERN\\
    Espl. des Particules 1\\
    1211 Meyrin, Switzerland
    \And
    Sünje Dallmeier-Tiessen\\
    CERN\\
    Espl. des Particules 1\\
    1211 Meyrin, Switzerland
}

\begin{document}
\maketitle

\begin{abstract}
Data and software citations are crucial for the transparency of research results and for the transmission of credit. But they are hard to track, because of the absence of a common citation standard. As a consequence, the FORCE11 recently proposed data and software citation principles as guidance for authors.\\
Zenodo is recognized for the implementation of DOIs for software on a large scale. The minting of complementary DOIs for the version and concept allows measuring the impact of dynamic software.\\
This article investigates characteristics of 5,456 citations to Zenodo data and software that were captured by the Asclepias Broker in January 2019. We analyzed the current state of data and software citation practices and the quality of software citation recommendations with regard to the impact of recent standardization efforts. Our findings prove that current citation practices and recommendations do not match proposed citation standards. We consequently suggest practical first steps towards the implementation of the software citation principles.

\end{abstract}


\section{Introduction}
Citations are recognized as one of the key incentives to make data and software available, as researchers receive credit for their efforts (\cite{fecher_reputation_2015}, \cite{huang_willing_2012}, \cite{wallis_if_2013}, \cite{tenopir_data_2011}).\\ 
However, the implementation of data and software citation capture as means of accreditation is challenging, as data and software citations tend to be idiosyncratic. Only little is known about how authors cite data and software in publications and how the tracking of data and software usage can be improved.\\
Empirical studies of data and software citation practices show a low coverage of data citation databases such as the Data Citation Index (\cite{peters_research_2016}, \cite{peters_zenodo_2017}, \cite{park_research_2019}) and a coexistence of direct software citation practices and citations to software related research publications  (\cite{li_challenges_2019}). Thus, tracking citations is challenging because of the great diversity of citable objects related to or representing software. Additionally, citable software papers gain attention (\cite{hong_software_2013}) and thus increase the diversity of targeted objects. This is an issue for tracking use impact, as citation discovery systems track citation counts to individual targeted objects, which often reflects only a fraction of real software usage. Because of this ``dilution of citations'' over many individual targets, citation counts in citation discovery services cannot reflect the full impact of software. Besides this citation dilution, recent text analysis studies have also demonstrated an inconsistency of data and software citation formats (\cite{yoon_how_2019}, \cite{li_software_2016}, \cite{li_how_2017}, \cite{hwang_software_2017}) and a practice of informal data and software mentions in the main text (\cite{park_informal_2018}, \cite{li_co-mention_2018}, \cite{howison_software_2016}, \cite{pan_assessing_2015}, \cite{pia_geant4_2009}). This increases the challenge of tracking data and software usage, as informal mentions remain invisible to citation discovery services. The absence of a widely adopted citation standard is stated as one of the major blockers of direct data and software citations (\cite{mooney_anatomy_2012}). To solve this, efforts towards the endorsement of citation principles are being made. The joint declaration of data citation principles (\cite{data_citation_synthesis_group_joint_2014}) covers purpose, function, and attribution of data citations, but does not provide specific implementation mechanisms. For software citations, the FORCE11 Software Citation Working Group published six software citation principles (\cite{smith_software_2016}) as a base for future concrete implementation plans.\\
However, despite all efforts, the tracking of the impact of data and software based on individual citation counts remains problematic.\\

In January 2019, the CERN-hosted general-purpose repository Zenodo announced the implementation of the Asclepias Broker (\cite{nielsen_software_2019}) together with the American Astronomical Society (AAS) and the NASA/SAO Astrophysics Data System (ADS). The Asclepias Broker is at its core a citation links store that is capable of harvesting discovery services such as the Astrophysics Data System, Crossref Event Data and Europe PMC for citation links via their open public APIs. The citation links found in the discovery services are either deposited (Crossref Event Data) or text-mined from the reference section (ADS and Europe PMC).\\
The tracking of citations to Zenodo objects (software, data, literature) is of special interest, as Zenodo is widely recognized for hosting software (\cite{brown_software_2018}) and because of Zenodo’s unique approach to versioning that enables aggregating citations to multiple individual objects. According to DataCite, Zenodo hosted more than 70\% of all persistently citable software in 2018 and thus became a pioneer and key player for the DOI registration for software (\cite{fenner_doi_2018}). Since 2014, software on GitHub can be integrated in Zenodo in order to retrieve a DOI (\cite{github_making_2016}). Zenodo’s unique approach of minting a DOI for each version of an object and an additional ``concept DOI'' representing all versions is especially suited for dynamic software that often have complex versioning schemes. This feature further facilitates tracking the impact of software, as citation counts to multiple individual objects can be aggregated using the concept DOI and its metadata links to the individual versions.\\
Thus, data on Zenodo’s citation counts is especially qualified for the exploration of the impact of software citation recommendations (i.e. an author-provided preferred citation) on actual citations.\\ 
This article explores the state of data and software citation practices by analyzing the 5,456 citation links to Zenodo resulting from the implementation of the Asclepias Broker. For the evaluation of the impact of recent software citation recommendation standardization efforts, we compare our findings to the quality of software citation recommendations.

\section{Results}
\subsection{The state of data and software citation practices}
Trackable data and software citations are not in common practice yet and we can only observe a fraction of real usage. We observe multiple issues that indicate how bad the state of data and software citations really is.

\begin{figure}
    \subfigure[dataset related DOIs]{\includegraphics[width=0.49\textwidth]{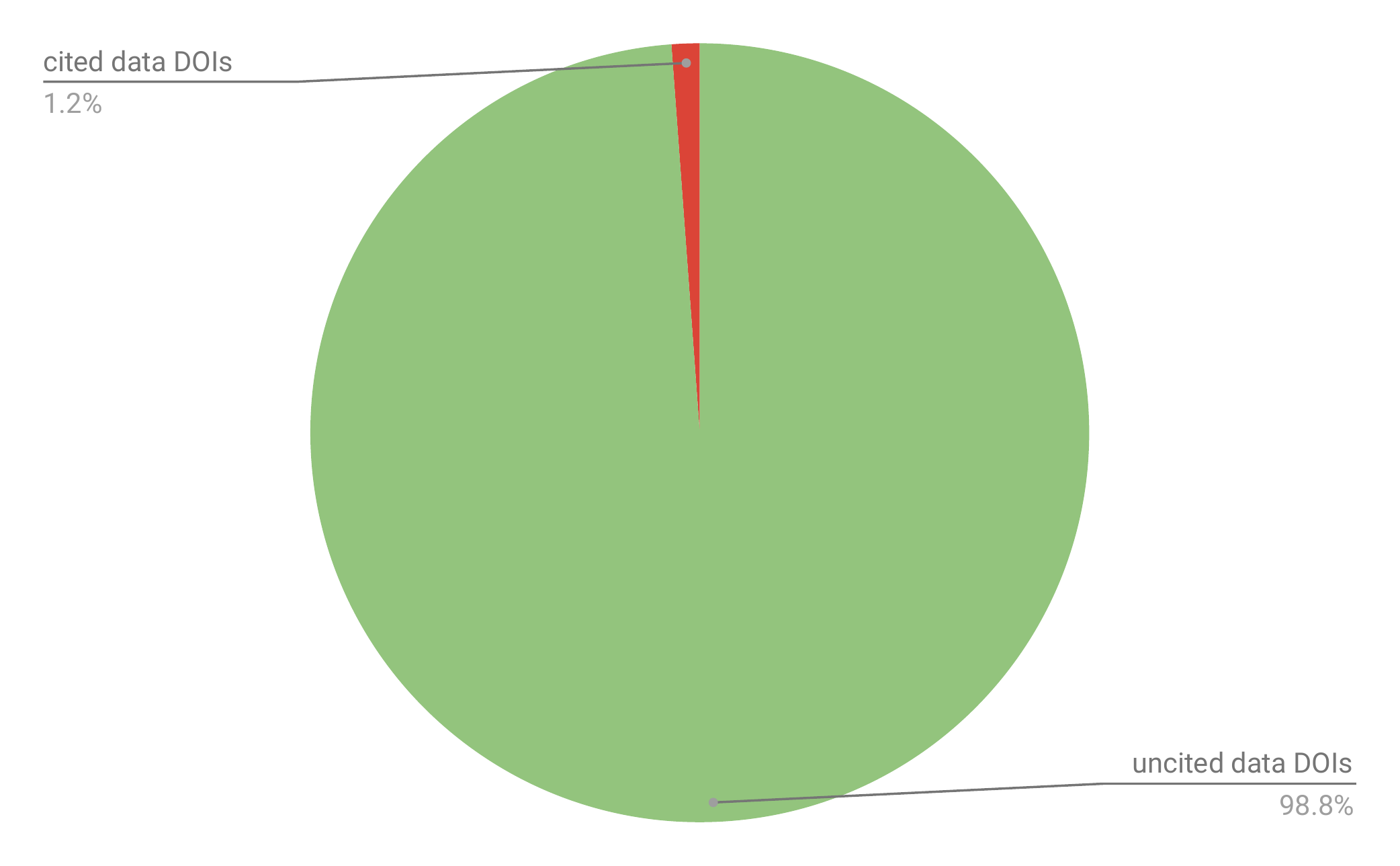}}
    \subfigure[software related DOIs]{\includegraphics[width=0.49\textwidth]{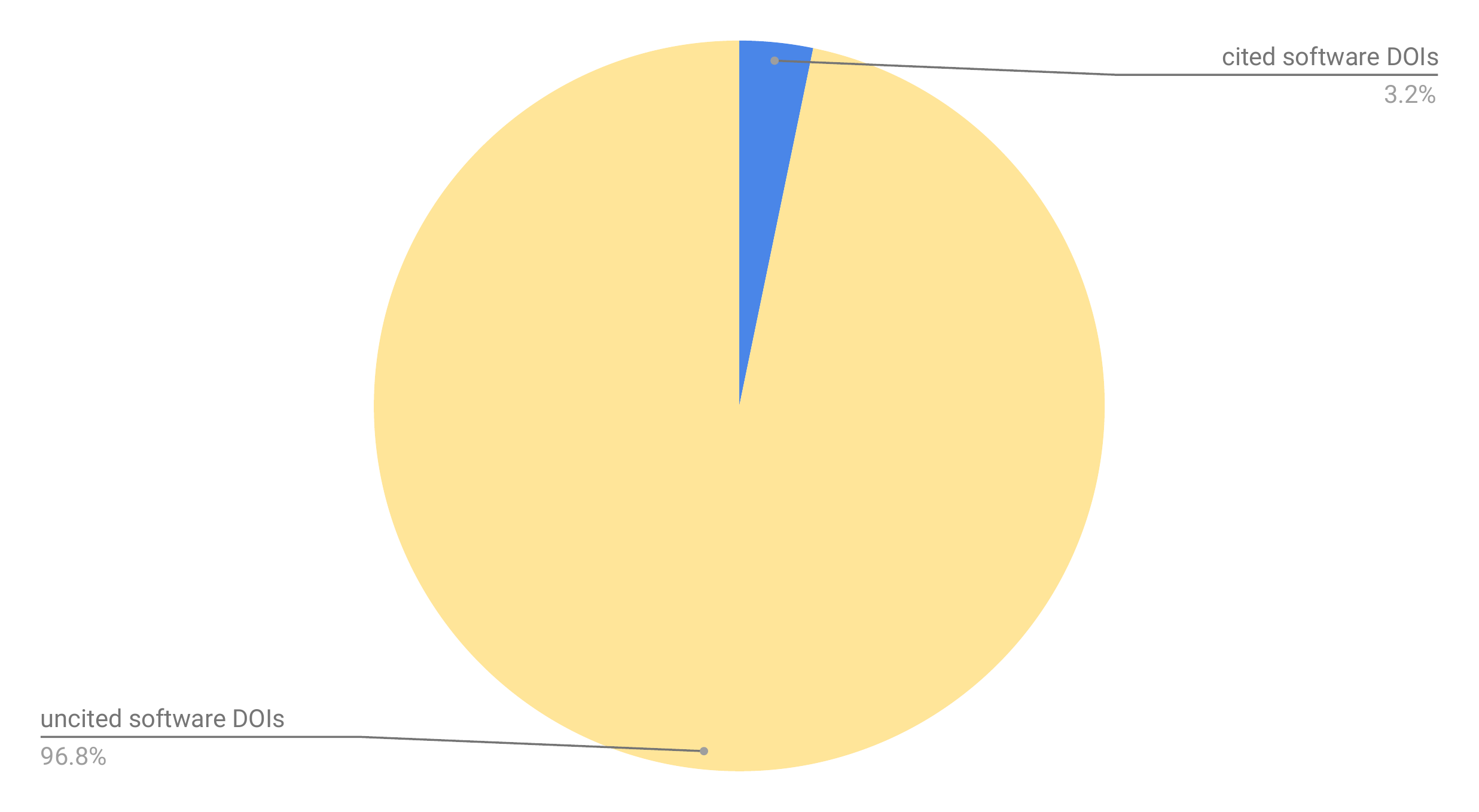}}
\caption{Citation coverage of dataset and software related DOIs}
\label{fig:12}
\end{figure}

\begin{figure}
    \centering
    \includegraphics[width=.6\textwidth]{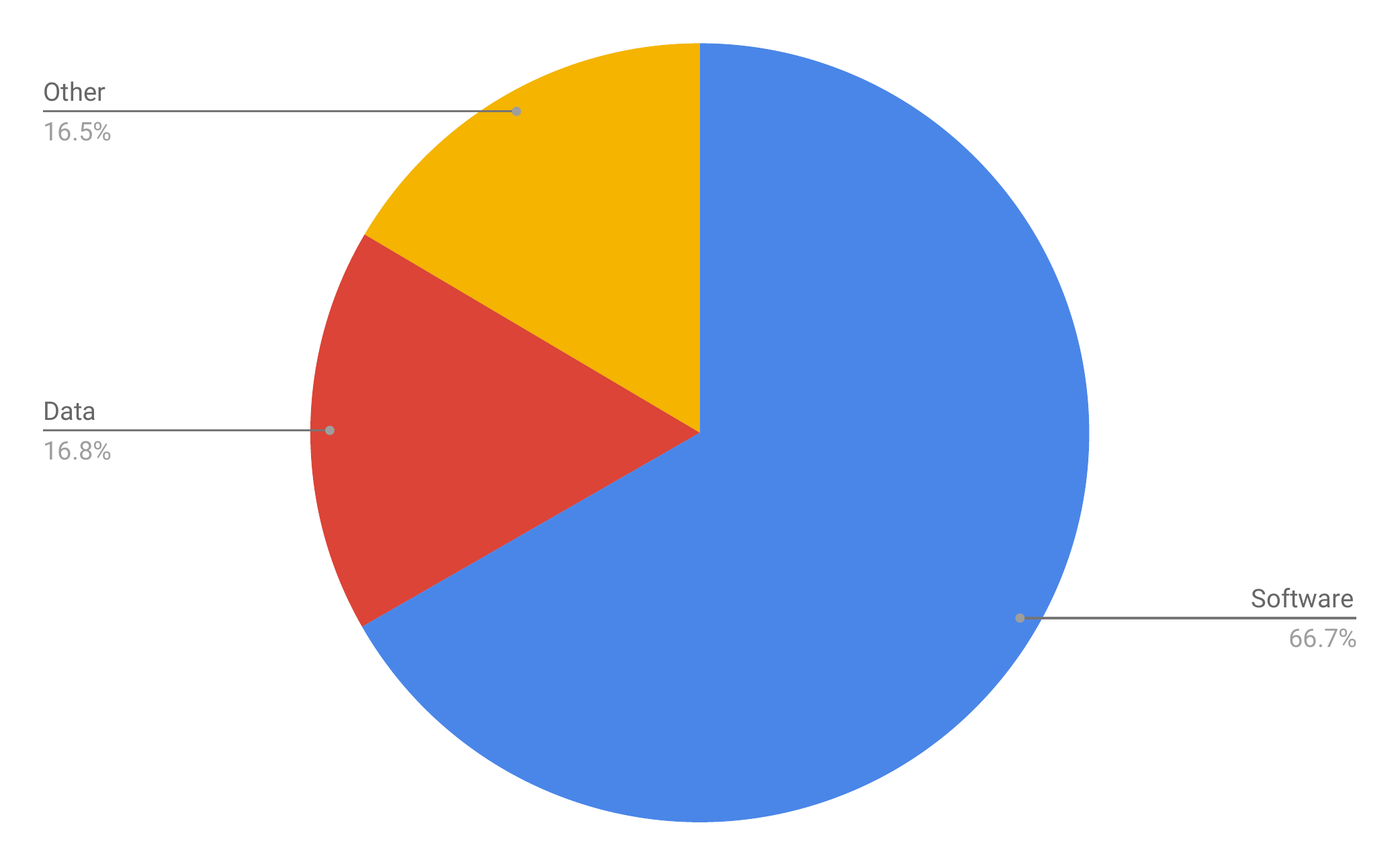}
    \caption{Type of cited object}
    \label{fig:3}
\end{figure}

\begin{figure}
    \centering
    \includegraphics[width=.6\textwidth]{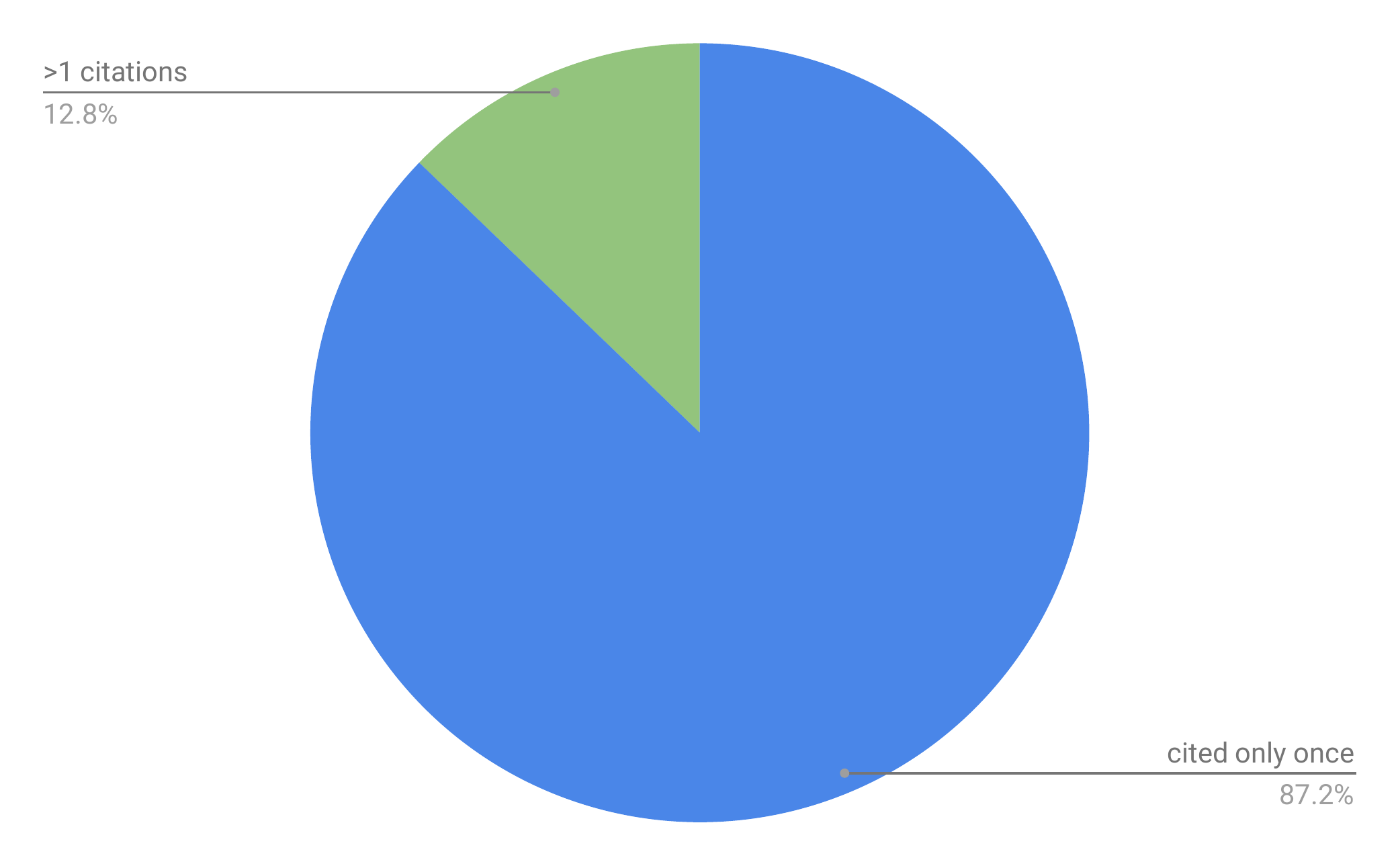}
    \caption{Citation intensity}
    \label{fig:4}
\end{figure}

\begin{figure}
    \centering
    \includegraphics[width=.6\textwidth]{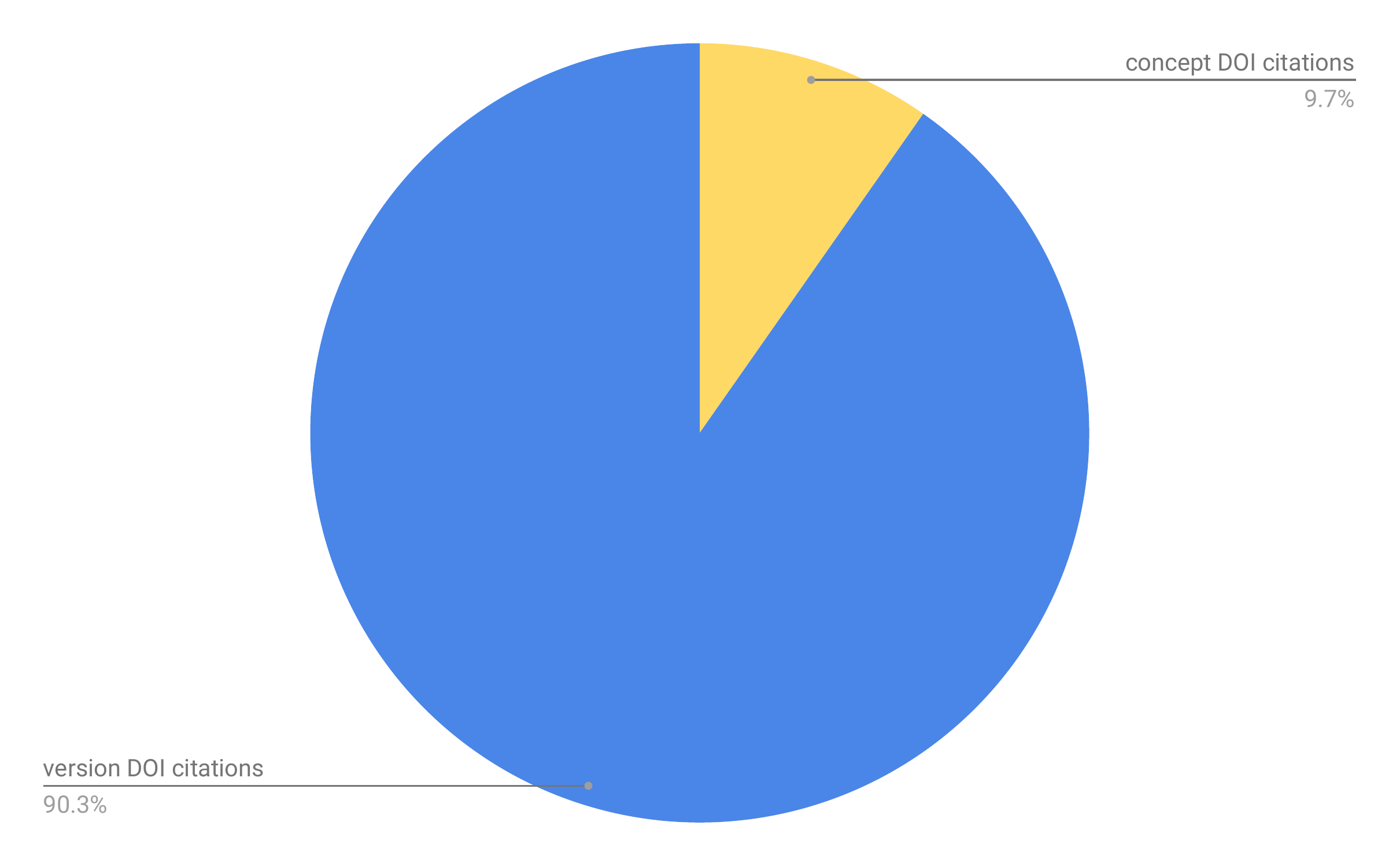}
    \caption{Concept DOI citation rate}
    \label{fig:5}
\end{figure}

We observe a very low citation coverage of Zenodo. Only 0.33\% of all DOIs, 1.16\% of dataset DOIs and 3.24\% of all software DOIs registered by Zenodo are traceably cited at least once (see Figure \ref{fig:12}). 66.7\% of all citations target software, while only 16.8\% target datasets and 16.6\% target literature (see Figure \ref{fig:3}). Besides a low citation coverage, we also observe a low citation intensity, as 87.2\% of all cited DOIs were only cited once (see Figure \ref{fig:4}). Only a few DOIs of broadly reusable software libraries such as LMfit (\cite{newville_lmfit-py_2019}) received high citation counts up to 97. The average citation count per cited DOI is 1.33. The majority of citations targeted version-specific DOIs, which is in compliance with the software citation principles (\cite{smith_software_2016}). The concept DOI was targeted by 8.31\% of all citations to objects with a concept DOI (see Figure \ref{fig:5}), despite the fact that the concept DOI is not very visible on Zenodo landing pages. 

\subsection{Most software and data citations in Zenodo are self-citations}
In order to understand the context and circumstances of data and software citations, we analyzed the citation speed and self-citation rate. The citation speed measures the time between the publication dates of the cited objects and the citing objects (i.e. how long does it take for an object to be cited).\\ 
We found varying citation speeds for different types of objects. For datasets the average citation speed is 0.45 years while for software it is 0.86 years. This means that datasets and software uploaded to Zenodo are likely to be cited within the first year after publication.\\
We estimate with 95\% confidence that 82.1\% of all citations with a citation speed between 0-1 years are self-citations (3,333 of 4,060 citations). The estimated self-citation rate for datasets is much higher compared to other types, as 98.5\% of all analyzed citations to datasets proved to be self-citations.

\subsection{Uniqueness of citation data sources}
\begin{figure}
    \centering
    \includegraphics[width=.9\textwidth]{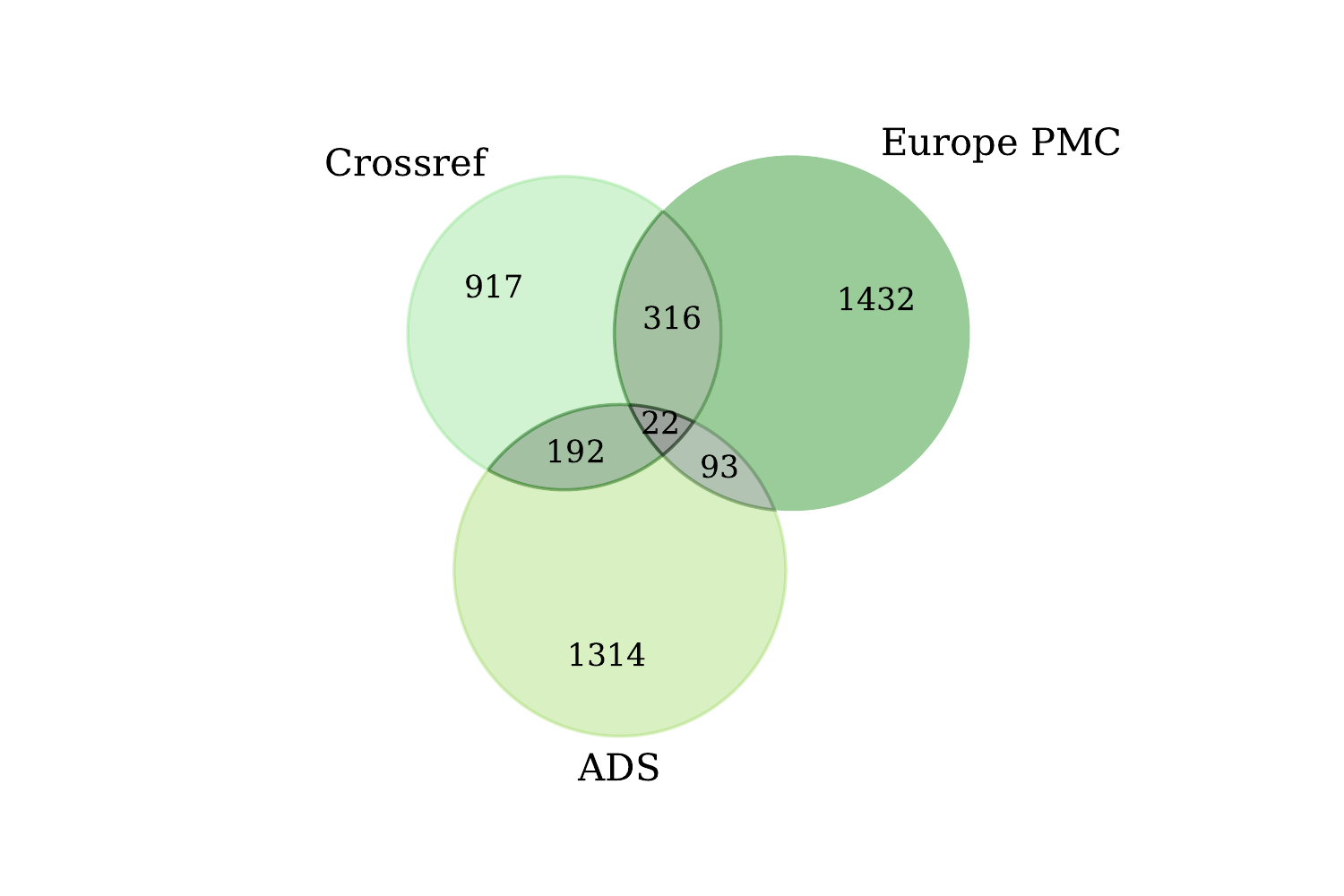}
    \caption{Intersection of citing literature discovered by different data providers}
    \label{fig:6}
\end{figure}

All three above observations in our data indicate that we can only track fractions of real data and software (re)usage. Issues with tracking citations significantly decrease the coverage of citation discovery providers.
It was certainly expected that each of the discovery services (ADS, Crossref Event Data, and Europe PMC) index citation links in content that is related or characteristic for the community they serve. But in fact, the overlap of citation links between services is so poor that we consider it almost non-existent, as demonstrated in Figure \ref{fig:6}. Unsurprisingly, only multi-disciplinary open access journals with open reference lists qualify to be captured by all discovery services. Thus, 18 of the 22 overlapping Zenodo data and software citing articles are published with Springer Nature’s ``Scientific Reports''. However, the real overlap may be estimated to be much bigger.

\subsection{The potential impact of software citation recommendations }
We analyzed the status and potential impact of citation recommendations (i.e. author-provided ``preferred citations'') for all 25 software projects archived on Zenodo that received 10 or more citations. Citation recommendations are meant to provide guidance and encourage citing software directly using persistent identifiers (PIDs). Therefore, it is interesting to see if the recommended identifier(s) were cited more frequently.\\

Recommended citations were often hard to find or when found outdated, confusing, or nonexistent. We frequently had difficulties finding what seemed to be ``hidden'' citation recommendations. Besides the GitHub ReadMe, software may be documented in various places: compiled documentation (e.g. ``Read the Docs''), PDF manuals, etc. Given that we had to invest some time to find information on how to give credit, these hidden citation recommendations are vulnerable to be missed by authors. It becomes especially complicated when developers change the name of the GitHub project or provide BibTeX snippets with contradictory DOI and URL metadata fields, as it happened in the case of ``triangle / corner.py'' (\cite{foreman-mackey_corner.py_2016}).\\

No trend towards changing recommendations in favor of citing the software directly via Zenodo DOIs was found. Once a research article is published, developers often recommend citing the article DOI instead of a Zenodo software DOI. In the case of ``SHTools'' (\cite{wieczorek_shtools_2019}), the developers listed both article and software DOIs as preferred citations for version 4.2, but subsequent releases dropped the direct software DOI from the citation block. On the other hand, some projects avoid dynamism as they recommend citation of an old version, e.g. ``LMfit'' (\cite{newville_lmfit-py_2019}). Six out of twenty-five analyzed projects did not provide a recommendation at all. The Zenodo DOI badge, which is a link the developer can add to  the GitHub ReadMe, was not counted as a recommended citation because it is not expressly declared as such and may easily be overseen or misinterpreted.\\

Mixed forms of recommendations propose more than one identifier or include contradictory statements. For example the software project ``Flavio'' (\cite{straub_flav-io/flavio:_2019}) recommends to cite an article DOI, but also highlights a Zenodo version DOI as a citable object.\\
Table \ref{tab:1} is a summary of our categorization of software citation recommendations. The total number of citations is the sum over citations to all versions and the concept DOI. The maximum age of the cited software measures the currentness of the cited object. A low number indicates that a recently published software version was cited. A high number indicates either that a potentially outdated version has been cited or that the software project was not updated for years. The diversity of a cited object expresses how many identifiers out of the total number of identifiers related to a software project were actually cited. The higher the count, the higher the dilution of citations across many individual objects.\\
Overall, there is little evidence that citation recommendations have an impact on the identifier choice. A more common phenomenon is that early version identifiers hardly ever become extinct, independent from any recommendation. Citation metadata in reference managers, once collected does not seem to be updated regularly and early adopters keep citing the first version identifier.\\

\begin{table}[]
\centering
\renewcommand{\arraystretch}{1.2}
\resizebox{\textwidth}{!}{%
\begin{tabular}{lllcccc}
\multicolumn{2}{l}{\multirow{2}{*}{\textbf{\begin{tabular}[c]{@{}l@{}}Recommendation category\\ \\ (August 2019)\end{tabular}}}} & \multirow{2}{*}{\textbf{Name}} & \multicolumn{1}{l}{\multirow{2}{*}{\textbf{\begin{tabular}[c]{@{}l@{}}Citation \\ count\end{tabular}}}} & \multicolumn{1}{l}{\multirow{2}{*}{\textbf{\begin{tabular}[c]{@{}l@{}}Max. age of \\ cited object\\ (in years)\end{tabular}}}} & \multicolumn{2}{c}{\textbf{\begin{tabular}[c]{@{}c@{}}Diversity of \\ cited object\end{tabular}}} \\
\multicolumn{2}{l}{} &  & \multicolumn{1}{l}{} & \multicolumn{1}{l}{} & \multicolumn{1}{l}{\begin{tabular}[c]{@{}l@{}}cited\\ DOIs\end{tabular}} & \multicolumn{1}{l}{\begin{tabular}[c]{@{}l@{}}existing\\ DOIs\end{tabular}} \\ \hline
\multicolumn{1}{l|}{\multirow{19}{*}{\textbf{\begin{tabular}[c]{@{}l@{}}Citation\\ recommendation\\ in place\end{tabular}}}} & \multirow{2}{*}{Publication} & GoaTools & 11 & 3 & 1 & 7 \\ \cline{3-7} 
\multicolumn{1}{l|}{} &  & SHTools & 10 & 4 & 4 & 12 \\ \cline{2-7} 
\multicolumn{1}{l|}{} & Software paper & Triangle/ Corner.py & 86 & 4 & 4 & 4 \\ \cline{2-7} 
\multicolumn{1}{l|}{} & Concept DOI & sncosmo & 10 & 4 & 1 & 4 \\ \cline{2-7} 
\multicolumn{1}{l|}{} & \multirow{7}{*}{\begin{tabular}[c]{@{}l@{}}Most recent\\ version DOI\end{tabular}} & Lasagne & 76 & 3 & 1 & 1 \\ \cline{3-7} 
\multicolumn{1}{l|}{} &  & Hyperspy & 25 & 2 & 9 & 25 \\ \cline{3-7} 
\multicolumn{1}{l|}{} &  & Python-Fsps & 17 & 4 & 1 & 1 \\ \cline{3-7} 
\multicolumn{1}{l|}{} &  & Pv & 15 & 4 & 2 & 8 \\ \cline{3-7} 
\multicolumn{1}{l|}{} &  & py-sphviewer & 10 & 3 & 1 & 1 \\ \cline{3-7} 
\multicolumn{1}{l|}{} &  & Firehose & 10 & 3 & 1 & 1 \\ \cline{3-7} 
\multicolumn{1}{l|}{} &  & Loprop for Dalton & 10 & 4 & 1 & 1 \\ \cline{2-7} 
\multicolumn{1}{l|}{} & Old version DOI & LMfit & 99 & 4 & 3 & 22 \\ \cline{2-7} 
\multicolumn{1}{l|}{} & \begin{tabular}[c]{@{}l@{}}The version\\ that was used\end{tabular} & Trackpy & 37 & 4 & 7 & 11 \\ \cline{2-7} 
\multicolumn{1}{l|}{} & \multirow{6}{*}{Mixed forms} & Matplot & 20 & 2 & 5 & 40 \\ \cline{3-7} 
\multicolumn{1}{l|}{} &  & Gatspy & 15 & 3 & 2 & 5 \\ \cline{3-7} 
\multicolumn{1}{l|}{} &  & MixSIAR & 14 & 2 & 2 & 7 \\ \cline{3-7} 
\multicolumn{1}{l|}{} &  & Pybinding & 13 & 2 & 2 & 10 \\ \cline{3-7} 
\multicolumn{1}{l|}{} &  & Flavio & 12 & 2 & 8 & 50 \\ \cline{3-7} 
\multicolumn{1}{l|}{} &  & NEST & 25 & 3 & 4 & 9 \\ \hline
\multicolumn{1}{l|}{\multirow{6}{*}{\textbf{\begin{tabular}[c]{@{}l@{}}No\\ recommendation\end{tabular}}}} & \multirow{3}{*}{\begin{tabular}[c]{@{}l@{}}DOI badge visible\\ on GitHub ReadMe\end{tabular}} & Seaborn & 88 & 4 & 6 & 9 \\ \cline{3-7} 
\multicolumn{1}{l|}{} &  & Plp & 20 & 4 & 3 & 10 \\ \cline{3-7} 
\multicolumn{1}{l|}{} &  & Cantera & 18 & 1 & 1 & 3 \\ \cline{2-7} 
\multicolumn{1}{l|}{} & \multirow{3}{*}{Nothing} & Hellwig & 18 & 3 & 1 & 1 \\ \cline{3-7} 
\multicolumn{1}{l|}{} &  & Uves\_Popler & 15 & 2 & 3 & 4 \\ \cline{3-7} 
\multicolumn{1}{l|}{} &  & \begin{tabular}[c]{@{}l@{}}Eddy-Covariance\\  Software TK3\end{tabular} & 11 & 3 & 1 & 1
\end{tabular}%
}
\caption{Software citation recommendation categorization}
\label{tab:1}
\end{table}

\begin{figure}
    \centering
    \includegraphics[width=.9\textwidth]{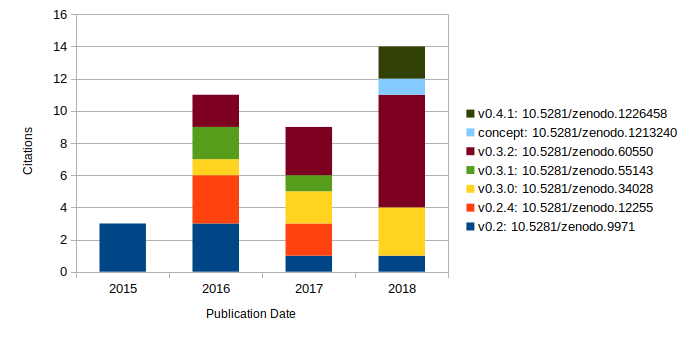}
    \caption{Diversity of cited DOIs for ``Trackpy'' (\cite{allan_trackpy_2019})}
    \label{fig:7}
\end{figure}

\begin{figure}
    \centering
    \includegraphics[width=.9\textwidth]{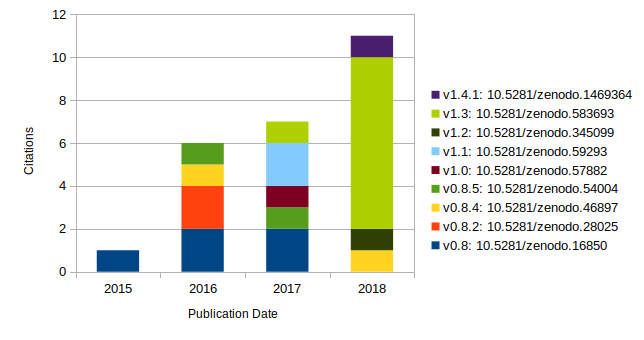}
    \caption{Diversity of cited DOIs for ``Hyperspy'' (\cite{pena_hyperspy_2019})}
    \label{fig:8}
\end{figure}

There are cases such as ``LMfit'', for which 97 of 99 citations were to the recommended but outdated version 0.8.0 DOI even after 4 years and the release of 22 additional versioned software DOIs. This impression is supported by Figure \ref{fig:7} and Figure \ref{fig:8}, which demonstrates how an identifier choice evolves over time for other packages. Even after many years and more recent versions to cite, the first version identifier keeps getting cited. We speculate that this latency is due more to static reference management systems (``BibTeX latency'') than it is to a continued (re)use of old software. This speculation is further justified when we find authors citing old software, but listing more recent (and citable) versions in the full text.

\section{Discussion}
Citation recommendations are meant to solve the problem of varying and thus hard to catch citation formats. But our results indicate that citation recommendations are in bad shape. They may even decrease the quality of software citations.\\
Software developers and researchers often do not receive guidance to provide high qualitative citation recommendations or how to use them. Despite the efforts of the FORCE11 to establish standardized software citation principles, software citation recommendations do not seem to change for the better. The software citation principles clearly recommend citing a PID that indicates which version of the software was actually used. In case there is an accompanying software paper, it may be cited alongside the version PID. 
But so far, the reality of software recommendations found in the wild greatly fails this ideal. Recommendations proved to be outdated, hidden, rather recommend to cite publications or do not even showcase a Zenodo DOI badge. Even if developers are enthusiastic enough to provide recommendations, it is tremendously hard to maintain them manually over the years.\\
In the end, this means that the right people will not receive credit for their work, because citations are either not trackable or citations are attributed to wrong versions with different author lists. Developer teams are just as dynamic as the software they are working on, thus citing the correct version ensures that the right team members receive credit instead of only e.g. the initial authors.\\

The low citation coverage of Zenodo and the detected high self-citation rate point out that the majority of data and software usage indications are still hidden in informal mentions and indirect citations to related publications. This also leads to developers not receiving credit for their work, because the lack of ironclad citations means that informal mentions are not trackable.\\

The obstacles of citing the specific software version in publications may still be too high to overcome, as authors may be overwhelmed with the variety of potential identifiers or lack experience in citing software directly. Reliable citation recommendations as guidance are crucial to help authors cite software directly and thus increase the number of trackable citations, which eventually leads to the right developers receiving credit.\\
Therefore, we suggest the following as a minimal first step to achieve reliable and trackable citations and ultimately give credit to the right people:

\begin{itemize}
    \item As a practical implementation, developers should recommend to cite one persistent identifier that refers to the software. This identifier ideally represents the whole software project, which is less prone to impermanence. This recommendation does not require maintenance effort on the developer site, but ensures that the software project is cited directly in a trackable way. 
\end{itemize}

This first step is not compliant with the software citation principles, and is prone to that the right people do not receive credit, but it achieves the purpose of making the citation easily trackable. As the next level, developers should recommend to cite the persistent identifier of the version that was used. Additionally, if the software is cited in a general context only, developers should recommend to cite the persistent identifier that represents the whole software project.\\

However, enhanced standards alone will not lead to better citation practices. Many players are involved in the dilution of citations. A citation evolves through a complex chain link system, which is extremely vulnerable to errors. Especially incorrect or ambiguous data and software citations are likely to be missed out, as the chain link system is not prepared to deal with non-traditional citations.\\ 

From the perspective of a discovery system, we observe that the technical presentation of data and software citations often lacks quality and precision, which complicates the parsing. During our analysis we have uncovered a number of issues associated with the formatting of citations in the source papers. Some of them indicate careless citation practices by the authors of the papers, while others suggest problems in the editorial supply chain system. Ambiguous citations that refer to more than one identifier cannot be assigned to more than one cited object. Incorrectly structured data delivered by editorial systems prevents the DOI in reference lists from being discovered. Also, we observed that the publication dates of citing and cited objects are not always trustworthy, as versioned objects were often associated with incorrect metadata. All these technical issues decrease the capture rate of each citation data provider, as many data and software citations are missed out. The discovered technical issues complicate further the associated social issues with software citation practices, which is beyond the limitations of our paper. Overall, this again leads to the right people not receiving credit.\\

Incorrect and ambiguous citations have the potential of being trackable and thus to highlight that credit has been given. However, this can only be achieved by rigorous editorial and typesetting efforts. But, publishing workflows are complicated and it is not trivial to reach out to every individual involved in the process. Dedicated training must be provided not only to editorial teams, but also to the publishing platform’s production teams and conversion teams, which are almost beyond reach. After the publication process, discovery systems are just as vulnerable to miss citations and as a final step, repositories may fail to provide correct citation metrics to the end users.\\
Thus, the overall picture is too complex to be changed by recommendations of any kind only. However, we must understand the full extent of the problem before we can apply changes to the whole chain link system, and before credit is ultimately given where long due.

\section{Methods}
\subsection{Citation links harvesting}
We harvested citation links from three discovery systems: the NASA/SAO Astrophysics Datasystem (ADS), Crossref Event Data and Europe PMC. Only citation links with a target DOI in the DOI prefix 10.5281 (Zenodo’s DOI prefix) were kept. Citation links were expressed in the Scholix Exchange Format as a citation relationship between two persistent identifiers (e.g. ``arXiv ID cites DOI'').\\
Both ADS and Europe PMC do parse the reference sections of publications to extract citation links. Europe PMC only extracts persistent identifiers from the full-text, but does not validate the persistent identifier leading to their data to contain invalid persistent identifiers (e.g ``10.5281/zenodo.''). ADS validates the persistent identifiers, which also allows them to detect e.g. URLs in the reference strings and match them to a DOI. The collection we used from Crossref Event Data depends on the reference lists that publishers decide to make available and self-deposit for journal articles.

\subsection{Deduplication of citation links}
We collected metadata about title, authors, publisher, and publication date from ADS, Crossref, DataCite, and Europe PMC for each persistent identifier in a citation link relationship. Most importantly we also collected alternate persistent identifiers. This allowed us to deduplicate citation links from different providers. For instance, one provider might express a citation link between a bibcode (an ADS identifier) and a DOI, while another provider might express the exact same relationship between a DOI and DOI.

\subsection{Cumulative citation counts}
We further annotated each deduplicated citation link with its target ``concept DOI'' if it existed in Zenodo. The ``concept DOI'' was obtained by querying Zenodo’s REST API (the concept DOI is also expressed in the DataCite metadata registered for Zenodo DOIs). This allowed us e.g. to investigate cumulative citation counts for multiple persistent identifiers for the same software project.

\subsection{Software}
The harvesting and deduplication of citation links explained above, was done using an installation at CERN of the Asclepias Broker (\cite{ioannidis_asclepias_2019}), which was developed as part of the Asclepias Project. The project is a Sloan-funded collaboration between American Astronomical Society, NASA/SAO Astrophysics Data System and CERN / Zenodo (further information can be found in (\cite{henneken_asclepias_2017})). The core of the Asclepias Broker is a store for citation links, the harvesting of citation links and metadata, as well as the deduplication and cumulative citation counts as explained in the above sections. The Asclepias Broker is open-source licensed under the MIT license.

\subsection{Dataset}
The dataset was retrieved from the Asclepias Broker early January 2019 after having performed a full harvesting and deduplication cycle from a clean database with zero citation links. The resulting 5,477 citation links were cleansed for our analysis. Misspelled DOI citations were corrected and the resulting duplicates deleted. 5,456 unique publication-to-software, publication-to-data, and publication-to-literature links remained. Each row of our dataset represents one unique link between a citing object and a Zenodo DOI.\\

The dataset contains information on the provenance of the citation links (field ``providers''). A link can either be captured by Europe PMC, ADS (NASA/SAO Astrophysics Data System), Crossref Event Data (crossref) or by a combination of these providers.\\
Generally, the dataset contains metadata about the cited object (the target object) and the citing object (the source object).\\

The citing object is identified by a ``source identifier", which is either a DOI or an arXiv ID. All of our citing objects are of the ``source type'' literature. Because every journal article or preprint can cite multiple Zenodo objects, an identifier can appear multiple times as more than one citation link was created. The ``source publisher'' contains information on the publisher of the citing article or preprint and the ``source publication year'' indicates the year the publication was published.\\
We made use of the re3data subject classification in order to classify the disciplinary content of the citing object (``source discipline''). The subject classification used at re3data offers four hierarchy levels of (sub)disciplines of Engineering Sciences, Humanities and Social Sciences, Life Sciences, and Natural Sciences. We used the second hierarchy level (14 categories) for our subject classification matching. The DOI prefix of the citing object indicates the journal an article was published in (e.g. ``10.1103/PhysRevD'' is exclusive for the journal Physical Review D). The description of the journal’s aim and scope was then used as based for our classification. arXiv preprints were classified using the arXiv subjects.\\

Equal information can be found on the cited object. The ``target doi''' contains the identifier of the targeted object of the citation, which is always a Zenodo DOI. The DOI may either be a version-specific DOI or a ``concept DOI'', which represents all versions of a project. The ``target type'' contains the type of the cited object. This value can either be a ``dataset'', ``literature'', or ``software''. The ``target title'' includes the name string of the cited object and the ``target publication year'' describes the publication year of the Zenodo object. The publication year is defined by the user and does not necessarily reflect the upload date on Zenodo. The ``target version id'' is a hash value that refers to the concept of the project. Each version DOI and the concept DOI of a project all refer to the same version identifier value. Citation counts that are spread across versions can be summed up using the version id value. The ``citation speed'' value was generated by subtracting the publication date of the citing object from the publication date of the cited object. This value indicates how many years it took for an object on Zenodo to get cited.\\

Additionally, we generated a second dataset with context information. We used a Python script to harvest metadata via the Zenodo REST API and via the GitHub REST API. This data was needed for the evaluation of potential correlations between citations and usage metrics. For every cited unique Zenodo DOI, we fetched basic metadata from Zenodo.\\
The ``cited doi'' value contains the DOI of the cited object. Every row of the dataset represents information on a cited object DOI (in contrast to the original dataset that is based on citation events instead of being object-centered). The ``concept doi'' is either empty or contains the concept DOI. Empty values mean that there is no concept DOI to be fetched due to the fact that the concept DOI was introduced later. The column ``citations'' contains the cumulated number of citations to a unique DOI. The number was generated by summing up the number of citation links that contained identical target DOIs. The ``resource type'' reflects the type of the cited object, which can be a ``dataset'', ``publication'', or ``literature''.\\

Usage metrics reflect how often an object may have been used. Zenodo delivers information on downloads and views: ``unique downloads'', ``unique views'', ``version downloads'',  ``version views'', ``version unique downloads'', ``version unique views''. Unique downloads and views count usage within one hour-sessions of a unique visitor ID only once. Views and download statistics can be aggregated to version usage using the concept ID. The version usage statistics can be narrowed down to unique views and downloads per visitor ID per one-hour session. Additionally, the ``volume'' and the ``version volume'' indicate the size of an object. Zenodo counts usage actions by humans or machines, but excludes double-clicks and robots.\\

For every cited DOI on Zenodo that is linked to a GitHub repository, we fetched metadata from GitHub. The ``github url'' contains the link to the GitHub project page. This value is empty for those Zenodo objects that do not link to GitHub.\\
The dataset contains the following GitHub related usage metrics: ``github downloads'', ``github forks'', ``github stars'', ``github subscribers'', and ``github watchers''. The variable GitHub downloads tells how many times a repository was downloaded. GitHub forks count the number of a repository’s copies made by users. GitHub stars describe how many users starred the repository. Subscribers count the number of subscriptions to conversations in issues, pull requests and team discussions of a GitHub repository. GitHub watchers are users who gets notified of activity in a repository, but are not collaborators.\\
Additionally, ``github languages'', ``github licenses'', and ``github topics'' contain information about the software. GitHub languages contain the (programming or markup) languages of a repository. A repository may contain one or several languages. Each value is separated by a comma. In case a license was assigned to the project, the value according to the GitHub license classification is given in ``github license''. A repository has only one or none licenses. The content of a GitHub project may be classified by the developer using topics. The content of the field is written in square brackets, which can be empty or contain one or multiple comma-separated values.\\

\subsection{Data analysis}
We calculated basic statistics for our dataset such as the citation coverage, the citation frequency, and the average citation count and the distribution of citations across types.\\
The citation coverage of Zenodo was calculated using the number of unique DOIs (4,108) that were cited in comparison to the total number of Zenodo DOIs registered at DataCite in June 2019 (1,236,264). We used the number of citations to each type (dataset, literature, software) for the distribution of citations across types. The citation intensity describes how often a Zenodo DOI was cited. The citation counts for each cited Zenodo DOI were calculated by summing up the number of rows that contain the same Zenodo DOI (the number of duplicates). Out of 4,108 Zenodo DOIs that were cited at least once, 3,582 DOIs were cited only once. The average citation count was calculated using the arithmetic mean of the accumulated citation counts. We used a Python script to harvest metadata from the Zenodo REST API and retrieved information on how many cited Zenodo DOIs are linked to a concept DOI. Out of 4,108 Zenodo DOIs, 2,395 had a value for the concept DOI. In 233 cases, the string of the cited Zenodo DOI was identical to the concept DOI, which means that the concept DOI was cited.\\

The self-citation rate was estimated on the base of the ``citation speed''.  The citation speed indicates how many years passed between the publication date of the cited object on Zenodo and the publication date of the citing object. This measure was calculated by subtracting the year of the publication date of the cited object from the publication date of the citing object (because citations generally look backward). Bigger numbers mean that more years have passed after the cited object was published on Zenodo. We assume that it takes time for third-parties to reuse datasets of software for different projects, as they have to fully understand what was done before and then need some time for a new analysis and for writing an article. Therefore, citations with a low citation speed qualify to be potential self-citation candidates. We selected all 4,060 rows with a citation speed with 0 or 1 (which means that the Zenodo object was cited in the same year or within a year after the publication on Zenodo) and analyzed a representative sample on their self-citation rate. The minimum size of a representative sample was determined on a 95\% confidence interval. In our case, the minimum sample size is 352. The authors of the citing articles of these 352 citations were manually compared to the authors of the cited object on Zenodo. In case of a self-citation, both author sets share at least one value. 289 out of 352 proved to be self-citations according to our definition. This value allows us to estimate with 95\% confidence on the whole population.\\

The correlation between citation rate and usage metrics were calculated using metadata retrieved from Zenodo and GitHub via their REST APIs.\\
We calculated the Spearman’s rank correlation coefficient for each usage metric and the number of citations to a unique Zenodo DOI. We chose Spearman’s rank correlation because it compares the rank values between two variables instead of assessing linear relationships. This fitted our sample well, because it contains many extreme values and outliers as few general purpose software libraries were highly cited.\\
As usage metrics, we chose the unique view and unique download counts provided by Zenodo, which are computed based on the COUNTER Project Code of Practice for Research Data (\cite{fenner_counter_2018}). This value is deduplicated by accumulating actions of a unique visitor within one-hour sessions. We calculated the coefficient using all cited objects and then run our calculations on every type (dataset, literature, software) separately.\\
We calculated the Spearman’s rank correlation coefficient for  the accumulated citation count of every cited Zenodo DOI that is linked to Github and usage statistics retrieved from GitHub: forks, stars, subscribers, and watchers.\\

The overlap of the indexes of every citation discovery service (ADS, Europe PMC), was visualized using a venn diagram. The set of each provider consists of the citing object’s identifier. Thus, the overlap of sets indicate how many citing articles are found by more than one citation discovery service.\\

For our analysis of the state of software citation recommendations, we accumulated the citation counts for every concept using the ``target version id'' hash value of our dataset. We considered all software concepts that had a citation count of 10 or higher. For these 25 software concepts, we categorized the state of their recommendations as it was presented to us in August 2019. We always considered the text on the ReadMe page of the project’s master branch and read every documentation we found. Precisely, we followed links given on GitHub and googled the software. For every citation recommendation we found, we additionally tracked the change history across every version release in order to make statements on their dynamism.\\
Because of the concept DOI and its metadata links to the individual versions on Zenodo, we were able to analyze the diversity of cited DOIs and observe the time course of every version DOI’s usage. The existing DOIs were counted on Zenodo. The count includes all version DOIs and the concept DOI. Manual effort was needed for the accumulation, as some projects have more than one concept DOI. Contrasting, we counted the number of actually cited DOIs.\\
The maximum age of the cited object indicates how old the cited software version was. 

\section*{Data Availability}
The unprocessed dataset that was retrieved from the Asclepias Broker in early January 2019 is publicly available on Zenodo (\cite{ioannidis_zenodo_2019}).\\
The processed dataset we used for our analysis and the dataset containing usage metrics are published on Zenodo as well (\cite{van_de_sandt_citations_2019}).

\section*{Code Availability}
The open source code of the Asclepias Broker is publicly available on Zenodo (\cite{ioannidis_asclepias_2019}).\\ The Python script used for the venn diagram and for fetching metadata from Zenodo and Github can be found on Zenodo (\cite{van_de_sandt_citations_2019-1}). 

\section*{Acknowledgement}
The Asclepias project is funded by the Alfred P. Sloan Foundation.\\

This work has been sponsored by the Wolfgang Gentner Programme of the German Federal Ministry of Education and Research (grant no. 05E15CHA).\\

Zenodo has received funding from the European Union's Horizon 2020 research and innovation programme under grant agreements (OpenAIRE-Advance) no. 777541 and (OpenAIRE-Connect) no. 731011

\section*{Competing interests statement}
The authors have no competing interests.

\bibliographystyle{agsm}  
\bibliography{literature}  






\end{document}